# Non-constrictive bead immobilization leading to decreased and uniform shear stress in microfluidic bead-based ELISA

K. Mitra[1*], B. Geiger[1], P. Chidambaram[1], A. Maharry[2], R. Xu[1], M. Tweedle[3]

[1]*The Ohio State University Department of Biomedical Engineering*

[2]*The Ohio State University Department of Electrical and Computer Engineering*

[3]*The Ohio State University Department of Radiology*

[*]*Present address Rice University Department of Bioengineering*

Microfluidic biosensors have been utilized for sensing a wide range of antigens using numerous configurations. Bead based microfluidic sensors have been a popular modality due to the plug and play nature of analyte choice and the favorable geometry of spherical sensor scaffolds. While constriction of beads against fluid flow remains a popular method to immobilize the sensor, it results in poor fluidic regimes and shear conditions around sensor beads that can affect sensor performance. We present an alternative means of sensor bead immobilization using poly-carbonate membrane. This system results in several orders of magnitude lower variance of flow radially around the sensor bead. Shear stress experienced by our non-constrictive immobilized bead was three orders of magnitude lower. We demonstrate ability to quantitatively sense EpCAM protein, a marker for cancer stem cells and operation under both far-red and green wavelengths with no auto-fluorescence.

## I. INTRODUCTION

Microfluidic devices have traditionally played important roles in rare sample handling for applications such as sequencing, rare cell isolation, cell sorting and in-vitro diagnostics (1-3). Multi-marker panel based diagnostics are especially amenable to microfluidic platforms as the automation dramatically increases throughput while conserving sample and reagent use. Systems biology approaches have delineated the importance of multi-marker analysis in discerning the pathology state and have significantly increased the efficacy of multi-marker panels. Li et al. recently demonstrated a systems biology informed process of marker panel selection using multi-reaction monitoring mass spectrometers as instruments to select for the marker cluster with the highest positive predictive value (4). Implementing such tests at feasible costs and viable time spans would require point-of-care tools.

Bead based assays within microfluidic devices have been extensively reviewed with several platforms undergoing clinical testing for cardiovascular and cancer based pathologies (5-6). A majority of bead based systems utilize constrictions and weirs to immobilize the solid phase bead while the mobile phase with target analytes is flown over (7-9). While this maximizes analyte-detector interaction and provides a spatial coding system for multiplexing, constriction severely hampers fluid flow across the bead. As Krüger et al. discussed, velocity of the mobile phase increases with a constriction based system (10). Koo and Kleinstreuer also pointed out that sudden bends and contractions in a channel lead to increased radial velocity (11). This increase in velocity reflects the continuity principle demonstrated in Equation 1 below.

$$V_{initial}A_{inlet} = V_{final}A_{outlet} \qquad (1)$$

$$V\text{: velocity of mobile phase; } A\text{: cross sectional area}$$

By this principle, as the cross-sectional area of the outlet decreases to hold the solid phase in place, the final velocity of the mobile phase must increase. This creates non-uniform flow. Increased velocity of the mobile phase means also increased shear stress across the solid phase as well as more turbulent flow of the mobile phase, neither of which is ideal for the detection system. These effects not only over-saturate certain portions of the solid phase and hamper detection, but also limit the flow so as to potentially shear the detection moiety off the bead.

To resolve this problem, a three-dimensional microfluidic device was conceived. This device would consist of two layers of microfluidics channels, separated by a porous membrane. The solid phase would be located at the end of the top channel, directly above the membrane. The membrane would allow flow of the mobile phase through the inlet channel into the outlet channel, while keeping the solid phase in place. Therefore, the cross-sectional area of the outlet channel would not decrease, leading to uniform flow velocity. This would also allow for uniform radial flow around the entire solid phase.

In the following sections, we outlay a device with capability to perform multi-marker detection without constriction to retain the solid phase. We also exploited the radial uniform flow across the solid phase to utilize the said phase as a scaffold for single cell proliferation assays. We hypothesized that a bi-layer microfluidic device with a polycarbonate membrane in the middle can effectively trap solid phase beads for sandwich ELISA based quantification of biomarkers. Two independent protein analytes, murine IgG and human EpCAM were used on chip, producing calibration curves and to ascertain limits of detection.

Fluorescent readout along the region of interest (solid bead) was obtained by pixel by pixel intensity readout to produce a score that was associated with a given concentration of antigen.

## II. METHODS

To ascertain two important flow profile characteristics, shear stress and flow velocity computational models were devised and tested in COMSOL (COMSOL Inc., MA) for both constriction chips and the proposed OncoFilter design. Minimal shear stress is preferred as it minimizes chances of shearing off antigen-antibody binding (12). Radially uniform flow velocity is essential to ensure uniform opportunity for antigen-antibody binding (13). Differences in expected flow characteristics served as the basis for device design and fabrication.

The microfluidic device called the OncoFilter uses standard soft lithography techniques to create microfluidic channels on polydimethylsiloxane (PDMS). Transparency masks were constructed using designs constructed on AutoCAD (Autodesk Inc., CA). Masks were utilized to produce silicon molds using optical contact lithography (Stanford Microfluidics Foundry, CA). Sylgard-184 PDMS (Dow Corning, MI) was mixed with a cross linking agent in a 10:1 mass ratio. This was poured over the silicon mold. After degassing, molds were baked at 70°C for 2 hours.

To bond dual layers of PDMS chip with the 12 micron pore polycarbonate membrane (Whatman, NJ), an APTES enhanced reactive oxygen plasma mediated bonding strategy was used. Aran et al. detailed such a strategy previously and utilized similar device to separate plasma from whole blood (14). A detailed schematic of a single channel of the OncoFilter chip is depicted in Figure 1 below.

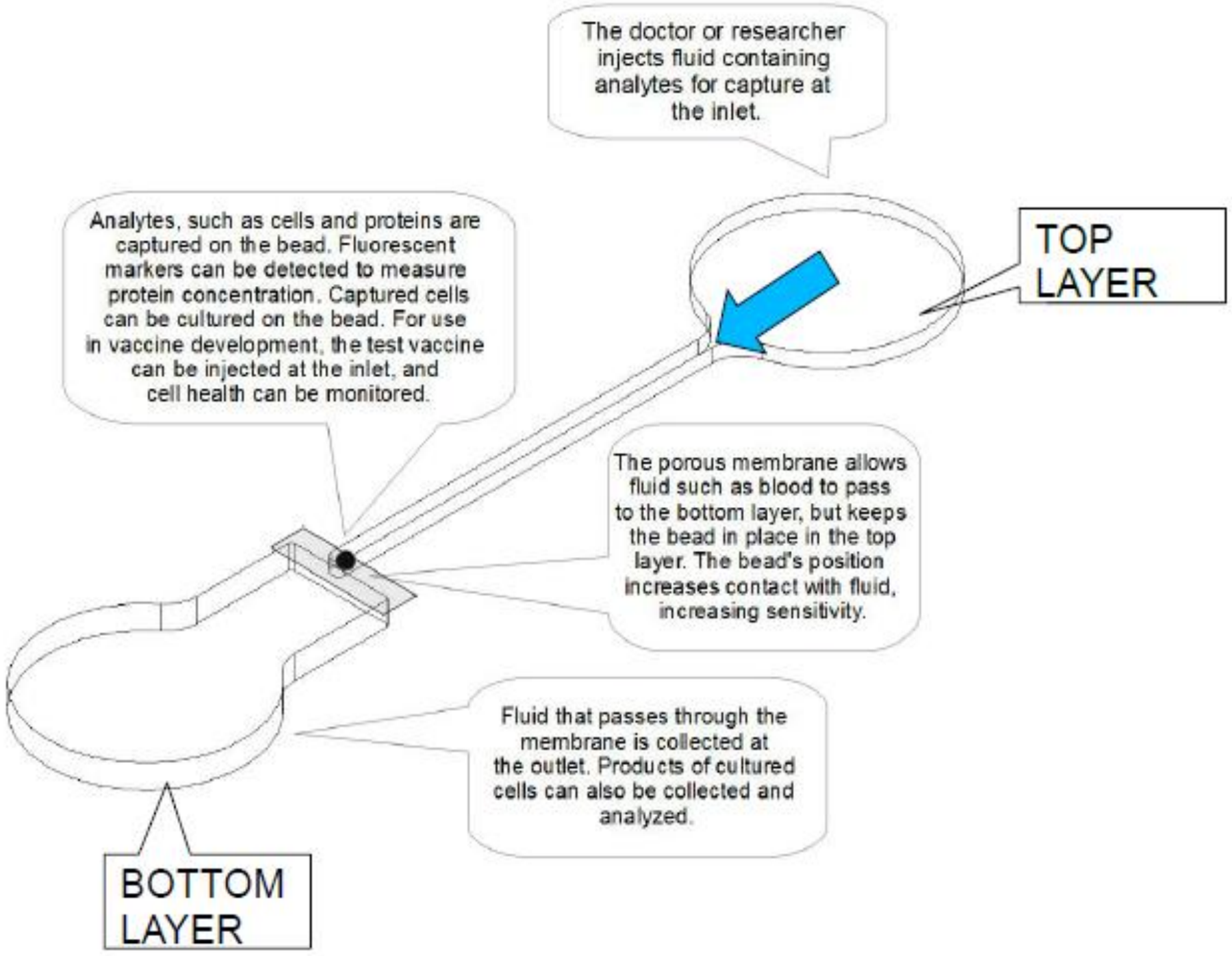

Figure 1: A single channel of the OncoFilter device. Multiple such channels are arrayed in parallel on a single device. The blue arrow indicates the direction of fluid flow.

Flow within channel was under constant observation of an optical microscope (Olympus, PA). Standard plastic tubing and 18 gauge needles were used to interface microfluidic channels with macroscopic flow systems. Luer-lock syringes were used as sample dispensing units with controlled release using a syringe pump.

Polystyrene beads (100 µm in diameter), which had high polydispersity (>99%) and were surface functionalized with streptavidin, were purchased from Spherotech Inc. The desired antibodies, either anti-EpCAM (Abcam, MA) or anti-IgG (Sigma Aldrich, MO), were purchased with biotin pre-attached to them. The streptavidin-biotin reaction conjugated the antibodies to the beads. Single beads were isolated from a 1% (by weight) solution of beads using a micropipette and an optical microscope. These were introduced into the microfluidic chip via the inlet and application of negative pressure.

After the antibodies captured the desired antigens, secondary antibodies attached to fluorescent markers attached to the other end of the antigens. These secondary antibodies were purchased and tagged with FITC fluorophores (for IgG) or IR-800 fluorophores (for EpCAM).

Antigen model EpCAM and murine IgG were utilized to demonstrate the ELISA microfluidic sandwich on the OncoFilter. The device was infused with bovine serum albumin and incubated for 30 minutes. To obtain a calibration curve, the antigen was diluted in 1 mL of phosphate buffered saline (pH 7.4) at predetermined concentrations. Flow was introduced into the chip at 1 mL/hr for 1 hour. Each chip housed five sensor beads, three which were exposed to flow and served as replicates and two which were unexposed to flow and served as controls. Each chip was flushed with PBS at 3 mL/hr for 30 minutes.

Intensity readout was performed using a homebrew image analysis system. Greyscale images of both sides of sensor beads were obtained via fluorescent microscope. Pixel by pixel intensity was calculated into a cumulative fluorescence score. Measured fluorescence was normalized to the maximum limit intensity.

## III. RESULTS

To ascertain flow parameters, a simple constriction immobilized sensor bead version of the OncoFilter chip was modeled in COMSOL. The shear stress and flow velocity profile as computed is presented in Figure 2 below.

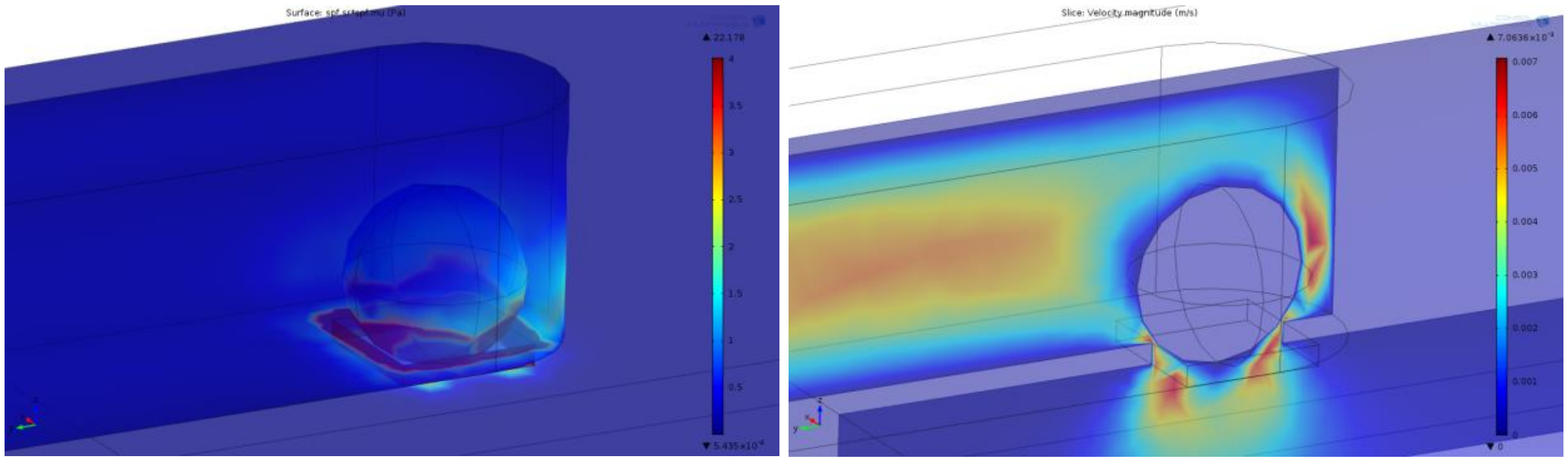

(a) (b)

Figure 2: Numerical model of the (a) shear stress and (b) velocity profile along the device midline for the constriction system. A region of high surface shear stress towards the base of the sensor bead interacting with the constriction trap. Shear stress in this region typically exceeds 20 Pa. The shear stress proved to be radially symmetrically distributed around the Z axis. The velocity for the fluid flow around the bead proved to be non-symmetric. A pincer effect towards the rear of the sensor bead drastically reduced cross-section of the flow area and increased flow velocity on average by a factor of 3.

In contrast, Figure 3 demonstrates the flow parameters obtained from a computational model of a non-constrictive membrane based bead immobilization in an otherwise identical fluidic system.

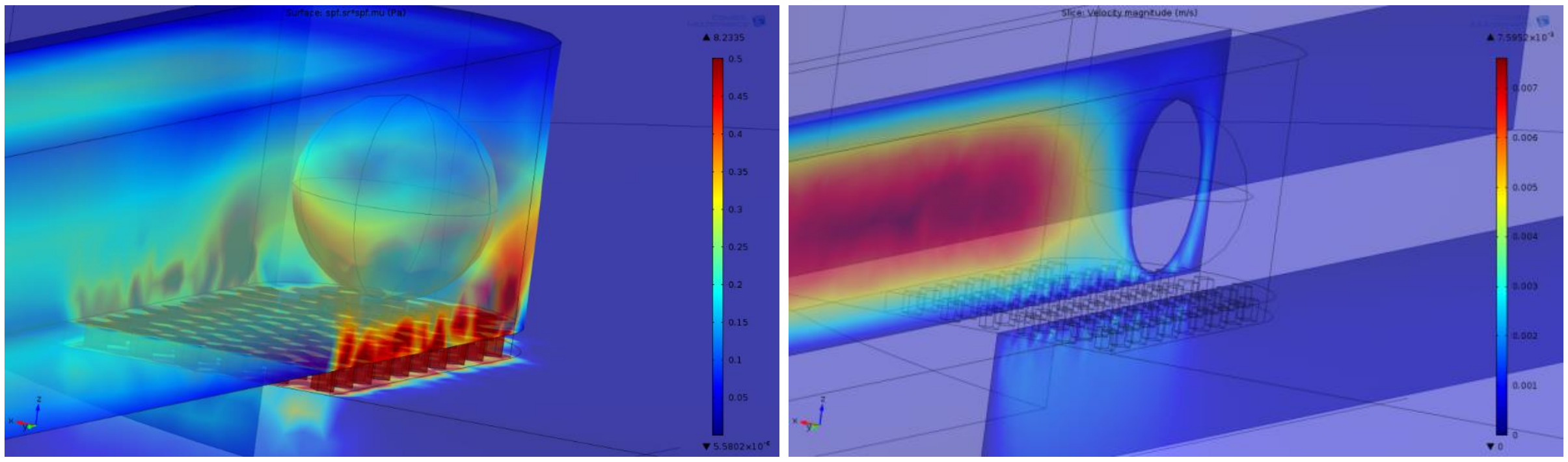

(a) (b)

Figure 3: Numerical model of the (a) shear stress and (b) velocity profile along the device midline for the non-constriction system. The maximum surface shear stress obtained from this model was 8.2335 Pa. The velocity for the fluid flow was shown to be symmetric.

The maximum surface shear stress obtained from the non-constrictive model of 8.2335 Pa was nearly a factor of 3 lower than the stress in the constrictive model. Also, it should be noted that it is unlikely the actual surface of the bead was subjected to the highest shear stress which was more likely experienced at the pores of the membrane. Importantly, the velocity of the flow profile was computed to be relatively uniform around the bead.

Velocity profile differences along the fluorescence measurement axis (in this case, the z axis), will cause differential exposure of the sensor bead to sample and concurrently, antigen. Figure 4 represents the models built to observe the difference in flow velocity along the measurement axis.

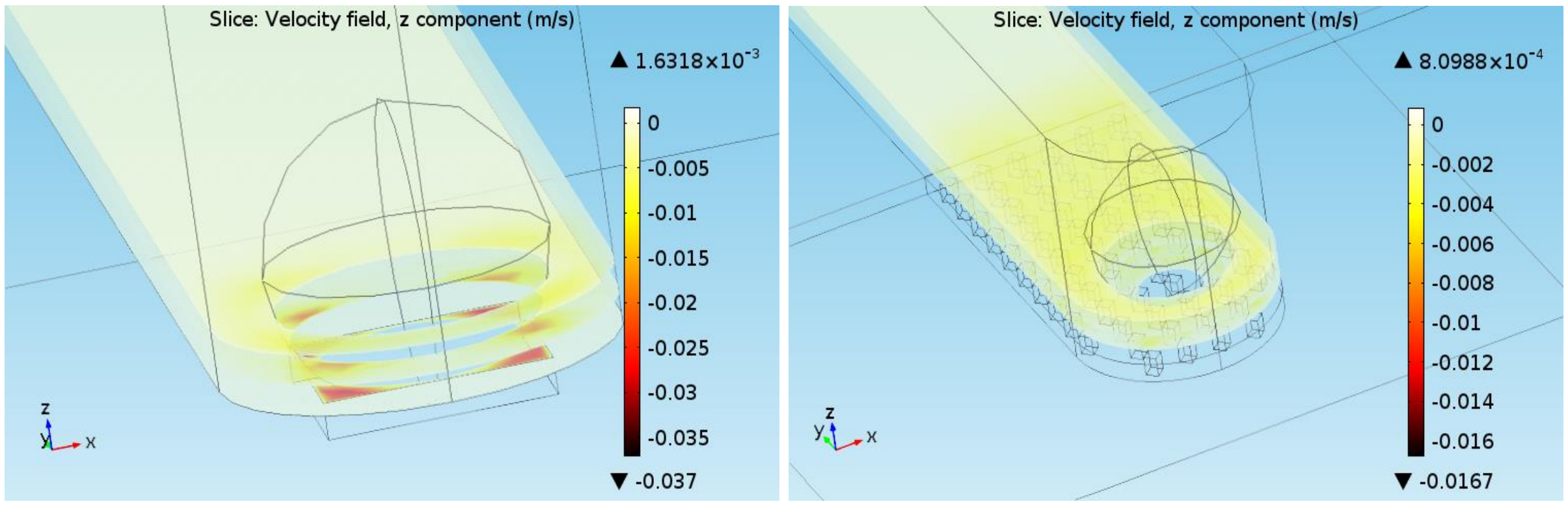

(a) (b)

Figure 4: Slice graphs of the velocity field in the z components at 10, 20, and 30 µm from the floor of the microfluidic channel for constriction (a) and non-constriction (b) systems.

The constriction version of the microfluidic chip exhibited regions of flow velocity in the bead proximity that were 6 to 7 fold higher than other regions likewise in bead proximity. Conversely, the velocity field of the non-constriction chip was computed to be uniform in the bead proximity, with the maximum velocity being half of that for the constriction chip. As Figures 5-7 show, when variation of flow along the circumference of the bead is plotted as a function arc angle, a clear difference is observed between constriction and non-constriction systems. Bottom, middle, and top layers were velocity slices 10, 20, and 30 µm from the bottom of the bead-microfluidic channel interface. Variability in velocity across arc angle was substantially higher in constriction systems with higher velocity at corners of the square constriction pits (40°, 130°, 220°, and 310°, which were 90° apart from each other). The variance in the velocity data for the constricted chip was also much greater than the variance of the porous chip data. For the bottom layer, the variance of the constricted chip was over 10 times greater than the variance of the porous chip. The constricted chip variance was almost 1000 times greater for the middle layer and almost 100 times greater for the top layer.

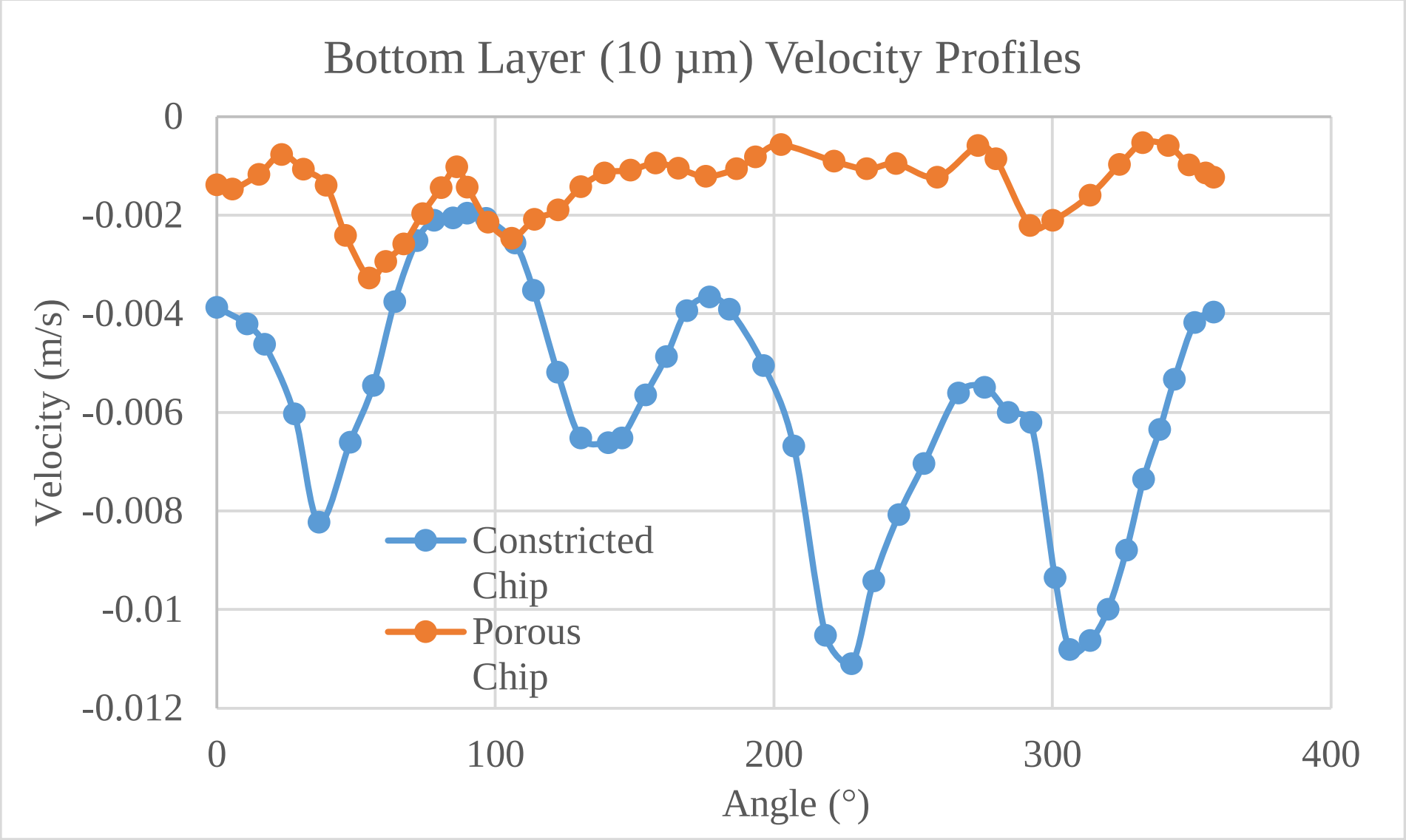

Figure 5: Velocity profile for the bottom layer (10 µm from the floor) of the constricted and porous chips. The constricted chip displayed a variance of 6.46x10$^{-6}$ (m/s)$^2$ and the porous chip displayed a variance of 4.41x10$^{-7}$ (m/s)$^2$.

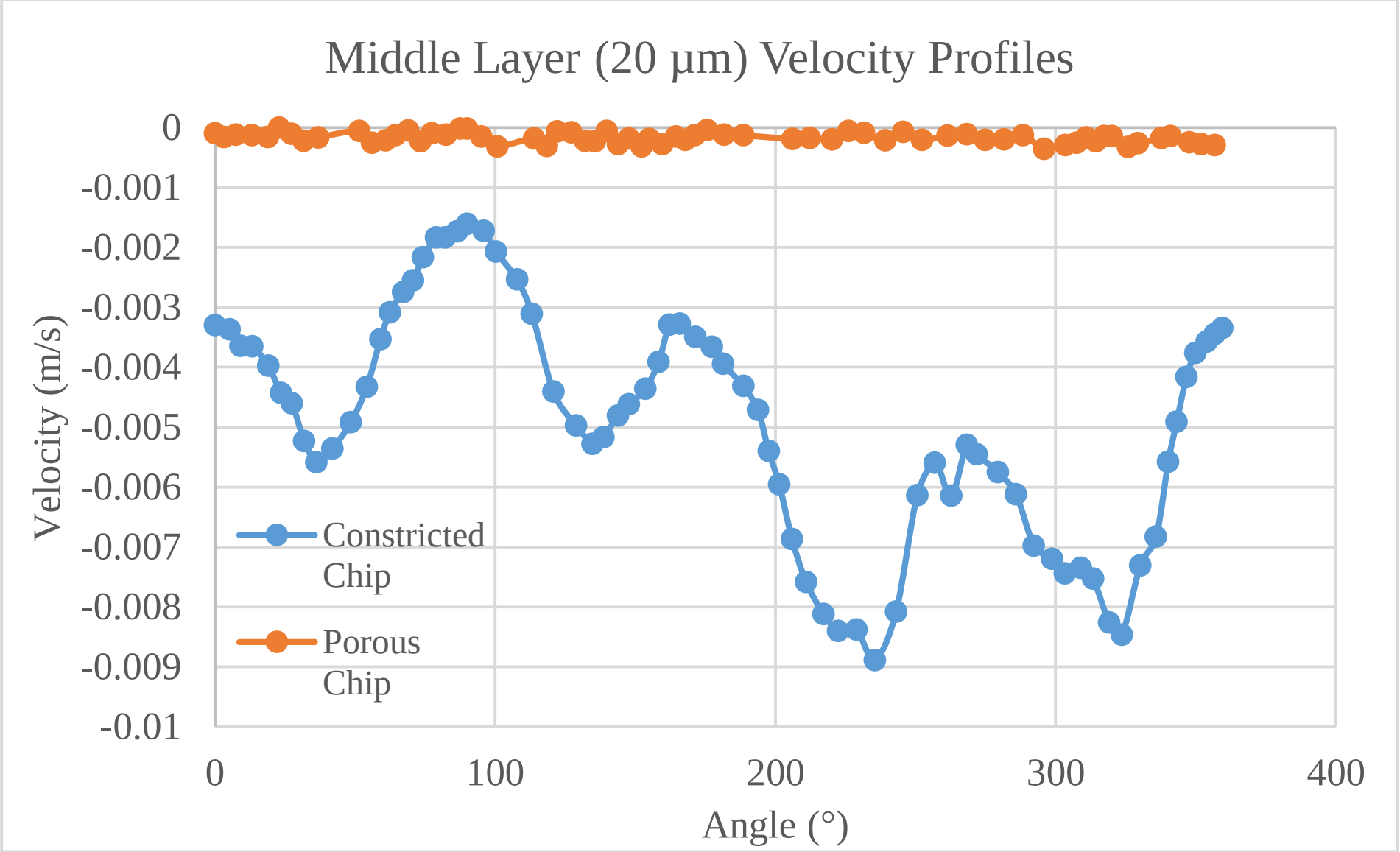

Figure 6: Velocity profile for the middle layer (20 µm from the floor) of the constricted and porous chips. The constricted chip displayed a variance of 3.69x10$^{-6}$ (m/s)$^2$ and the porous chip displayed a variance of 7.07x10$^{-9}$ (m/s)$^2$.

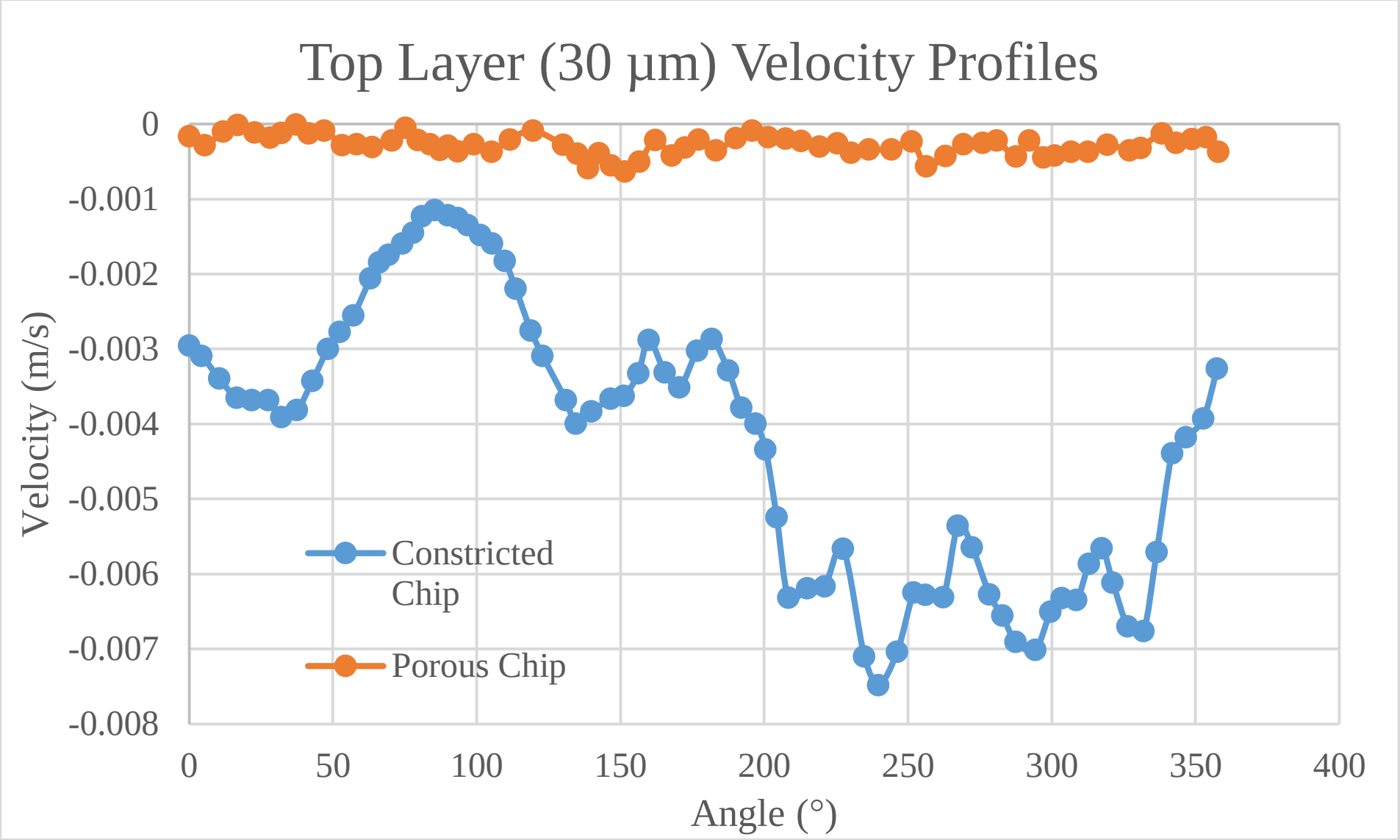

Figure 7: Velocity profile for the top layer (30 µm from the floor) of the constricted and porous chips. The constricted chip displayed a variance of $3.40 \times 10^{-6}$ $(m/s)^2$ and the porous chip displayed a variance of $1.83 \times 10^{-8}$ $(m/s)^2$.

Another advantage of a non-constriction based system is the ability to decouple the shape of the microfluidic conduit from the need to constrict the sensor bead against fluid flow. This allows choice of conduit geometries that allows for more even flow velocity along sensor beads. Figure 8 below computationally simulates this advantage.

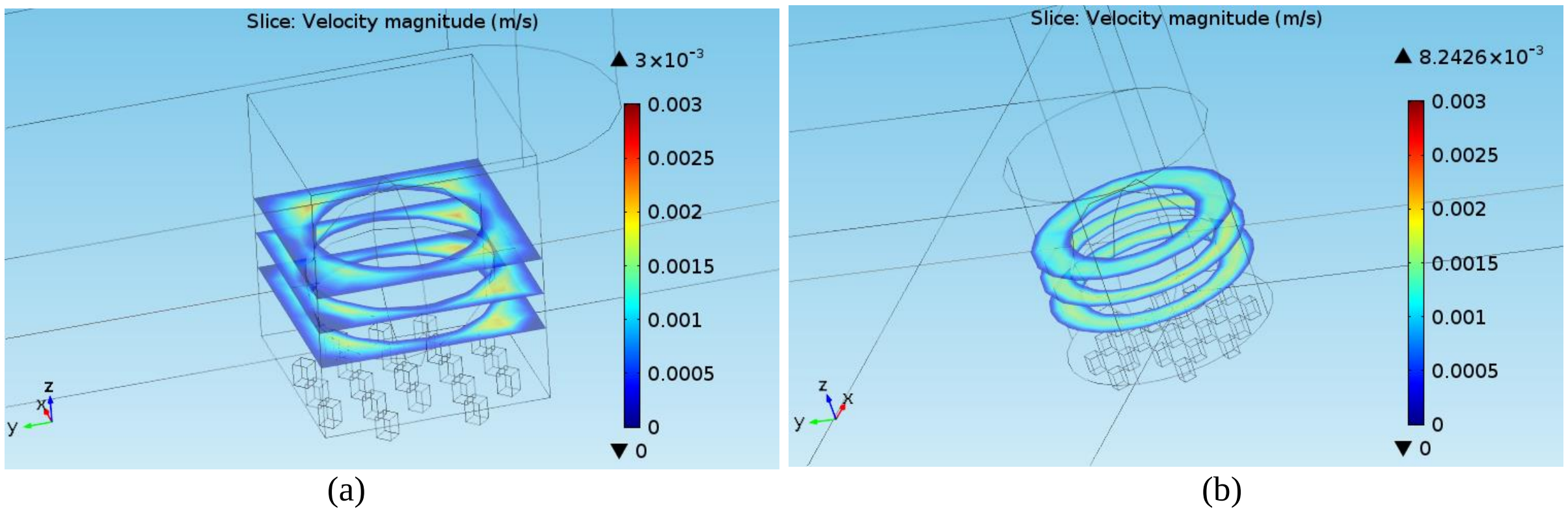

Figure 8: Flow velocity around a non-constriction chip with (a) a cubical conduit and (b) a cylindrical conduit.

To ascertain the feasibility of the non-constriction based device, we fabricated the chip shown in Figure 9 as designed and detected protein binding using IR800 CW fluorophore in a sandwich ELISA configuration.

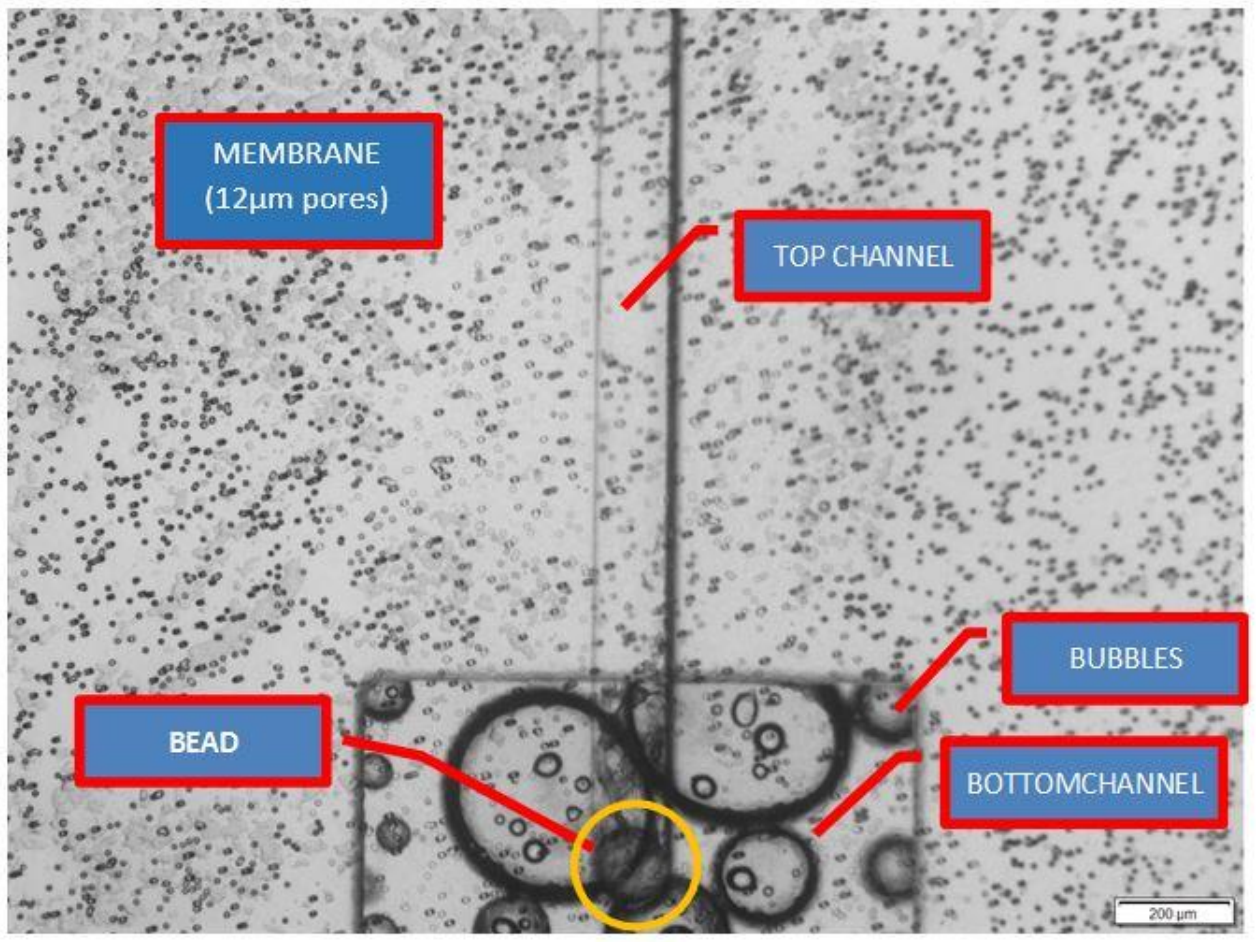

(a)

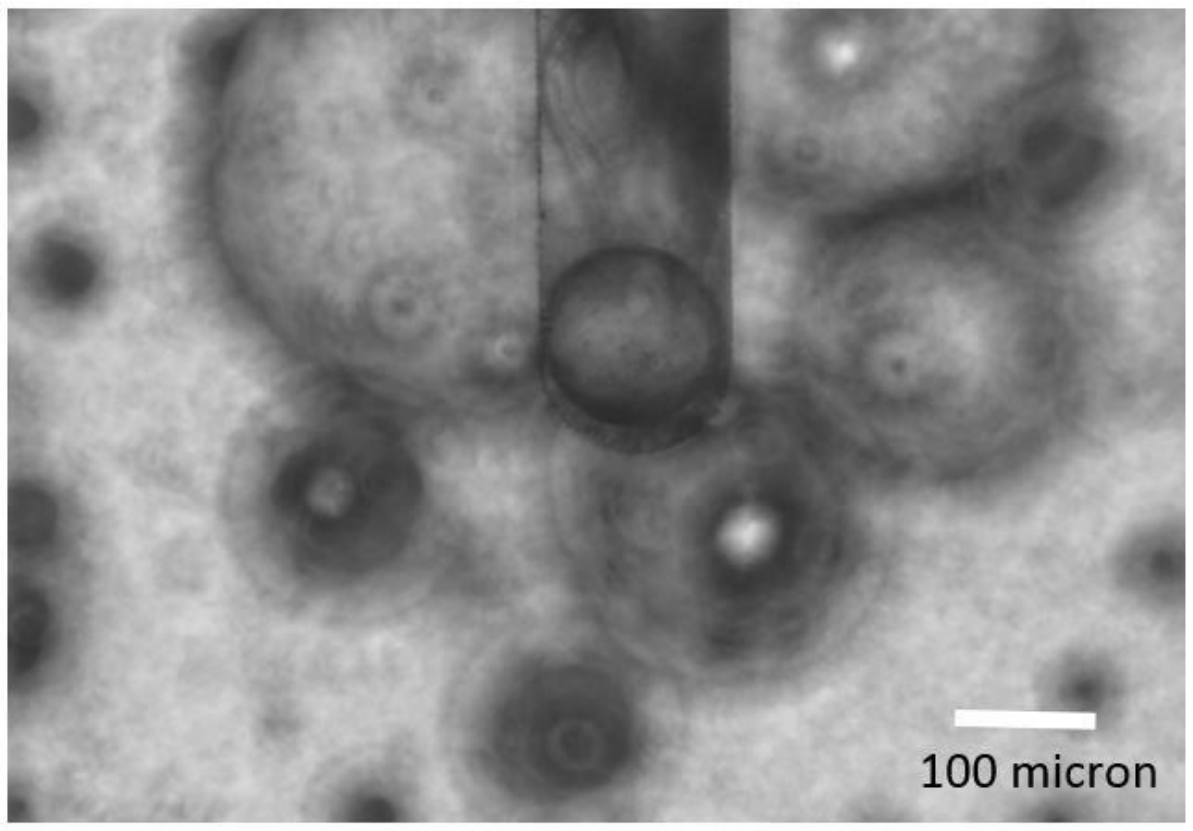

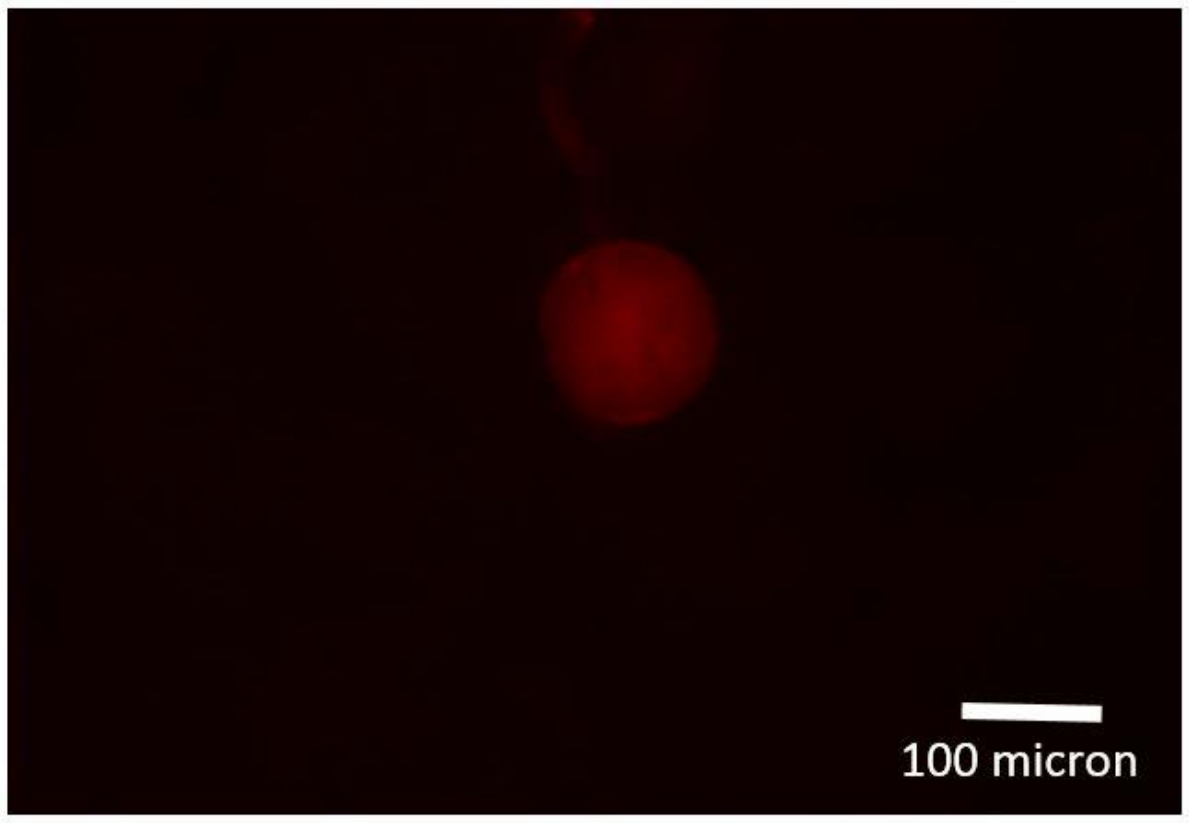

(b) (c)

Figure 9: (a) Sensor bead (circled in yellow) trapped within the microfluidic channel by the membrane. Bubbles can be removed by subjecting the system to a greater duration of flow and thus infusing the entire chip with fluid. (b) Bead trapped over the overlap region and over the membrane within the chip. Bead is conjugated with anti-EpCAM. (c) After infusion of EpCAM and secondary anti-EpCAM conjugated with IR-800 CW, chip is imaged under fluorescence microscope with IR 800 filter.

Intensity readout was performed using homebrew image analysis system. Greyscale images of both sides of sensor beads were obtained via fluorescent microscope. Pixel by pixel intensity was calculated into a cumulative fluorescence score. Maximum fluorescence concentration readout limit was set at 10 ng/mL for EpCAM. All subsequent measured fluorescence was normalized to the maximum limit intensity. The calibration curve obtained was plotted in Figure 10 below. The curve was found to be consistent to the expected log-linear trend.

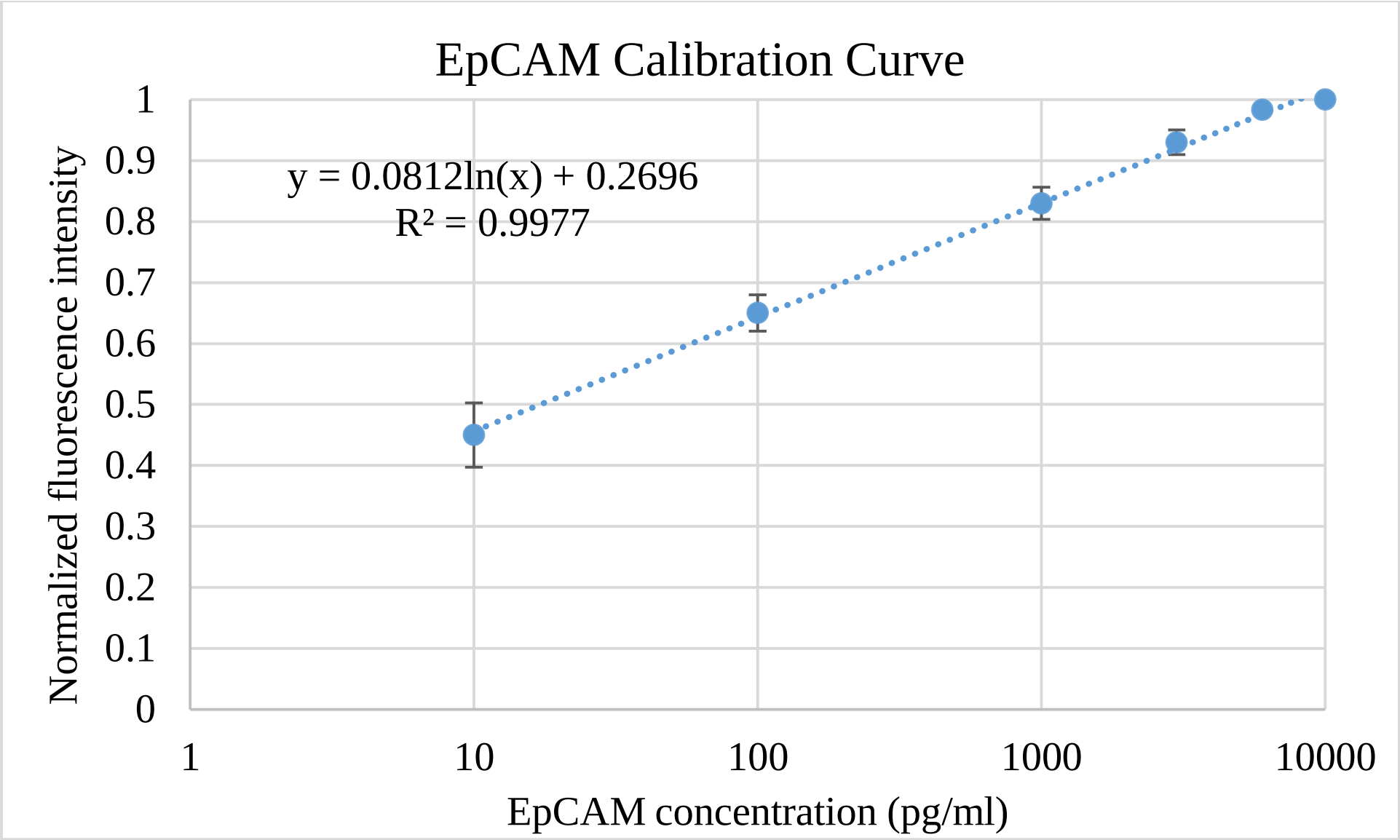

Figure 10: Calibration curve for varying EpCAM concentrations. Curve is plotted against fluorescence intensity of the bead measured by CCD imaging. Intensity readings are normalized to the fluorescence measurement obtained at 10,000 pg/mL.

To ensure the OncoFilter system was capable of detecting multiple antigen types and function with different fluorophores with minimal noise (background autofluorescence) and high specificity, detection tests were run using murine IgG and human EpCAM using FITC and IR-800 as reporters respectively. Qualitative comparison of FITC-based detection of murine IgG to the IR-800-based detection of EpCAM is depicted in Figure 11 below and shows minimal background signal and high specificity after processing of both bead with identical image processing algorithm.

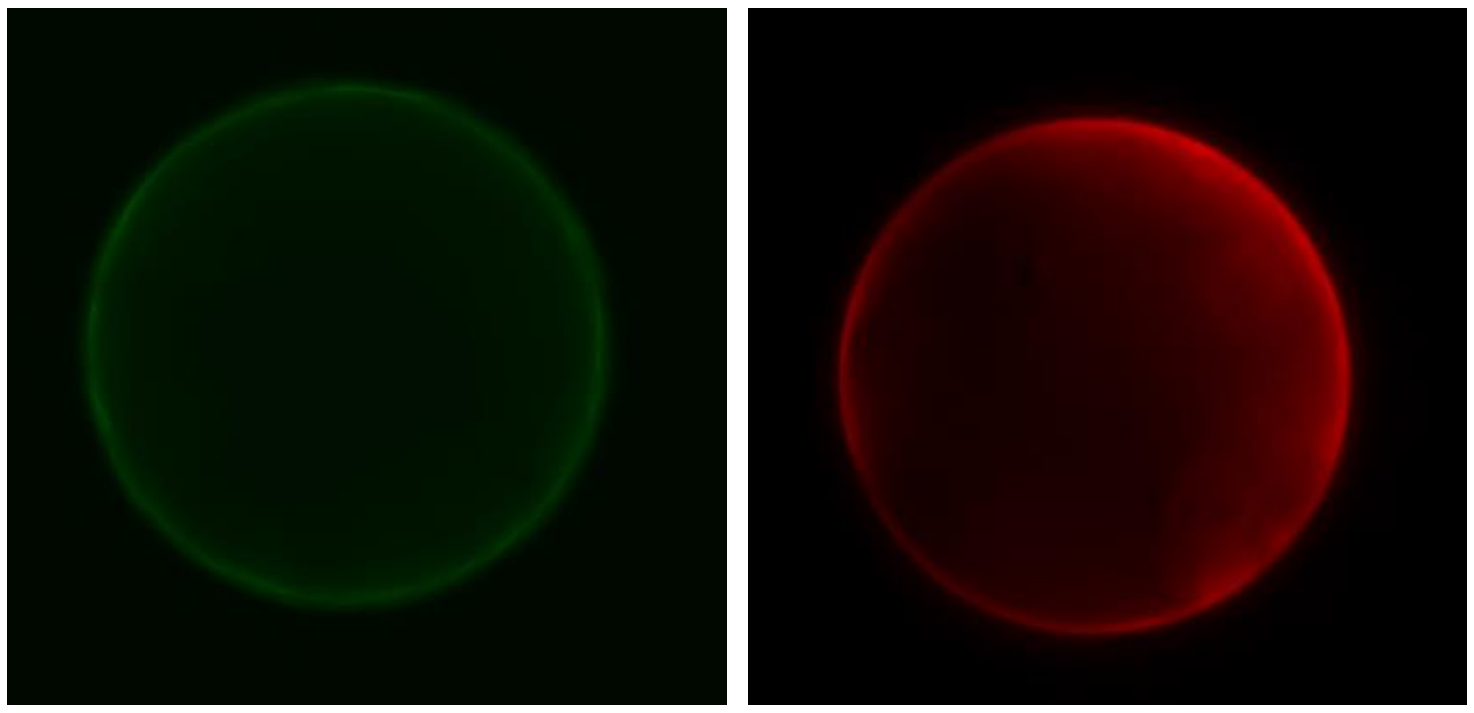

(a) (b)

Figure 11: Sensor beads that detects IgG using FITC (green) and EpCAM using IR-800 (red). Both sensor beads experienced uniform radial exposure to both antigens. Antigen exposure was maximal at bead margins. The higher quantum yield of FITC is evidenced by the slight background scatter around the bead. Comparatively, the IR-800 bead shows a higher signal-to-noise ratio though IR-800 dyes shows significantly lower fluorescent quantum yields.

## IV. DISCUSSION

Our work computationally demonstrates flow limitations of bead constriction based biochips, a common platform for ELISA. We model a non-constrictive bead based assay that demonstrates shear stress 3 times lower than that of the constriction based chip. Consistently, the flow velocity magnitude of the constriction based chip proved to be 3 times higher than that of the non-constriction based chip. Importantly, non-uniform flow profiles was observed on both YZ and XY planes of flow. Physically these planes represent the plane parallel to direction of flow and the plane perpendicular to flow respectively. Non-uniformity of flow along these planes in positions proximal to the bead would result in non-uniform exposure of antigen to the bead. The use of non-constrictive membrane based trapping of sensor bead resulted in one to three orders of decrease in variance of fluid velocity along the sensor bead apart from an overall reduction in the velocity of the flow. This would physically translate to more uniform distribution of antigen across the bead surface as well as lower shear force (proportional to the factor of decrease of velocity) on the antibody-antigen complex. The use of non-constriction based system also allowed for the use of conduit geometries that further mitigated non-uniform radial flow velocities. The device proved to be sensitive to antigens to the limit of 10 pg/mL of sample and sandwich ELISA protocol ensured high specificity in detection with a dynamic range spanning 3 orders of magnitude.

A significant challenge of device function was the occasional obstruction of the top flow channel by particulate debris that proved to be too large to exit via membrane pores. While the issue could be resolved by sample filtering, this option may not be always available especially in cases of ultra-sensitive disease testing. The membrane also experienced minimal wicking of fluid and bubbling from the exit of air trapped during the bonding protocol. Saturation of the chip with fluid by degassing proved an effective remedy. Further testing of this device will ensure optimization for specific applications and useful advantages of the non-constriction system.

## V. CONCLUSION

We present a microfluidic device for bead based ELISA that ensures radially symmetric flow around the sensor bead at low shear stress environment. We computationally evaluate the efficacy of the device to provide radially symmetrical flow and low shear stress flow conditions. We experimentally tested the ability of the device to quantitatively sense multiple antigens using fluorescent reporters of different wavelengths. In doing so we demonstrate variance change and F statistic.

We also find shear stress for the non-constriction device was at least 3 times lower than that for the constriction device. Calibration for detection of human EpCAM protein detection yielded a dynamic range of detection spanning 3 orders of magnitude. Detection of protein could be achieved at concentrations at least as low as 10 pg/mL of PBS. Our device was functional for multiple fluorophores demonstrating usability as a spectrally multiplexed diagnostic chip.

## ACKNOWLEDGEMENTS

The authors would like to thank staff at the Tweedle lab and the Ohio State biomedical engineering department for lab advice and imaging support. Thanks are also due to the staff at the Stanford microfluidics foundry. This work was made possible in part by the Pelotonia cancer fellowship and Venturewell E-Team grant.